\documentstyle[12pt]{article}

\begin{document}

\large

\centerline{\Large \bf Interaction of Dual N=1, D=10 Supergravity }
\bigskip
\centerline{\Large \bf
with Yang-Mills Matter Multiplet \footnote{Research supported by the
ISF Grant MOY000}}

\vspace*{ 1cm}

\centerline{\bf N.A.Saulina $ ^{\it a,b}$,
M.V.Terentiev $ ^{\it a}$,  K.N.Zyablyuk $ ^{\it a,c}$ }

 \bigskip

\centerline{$ \it ^{ a}\ Institute \ \ of \ \ Theoretical \ \
 and \ \  Experimental \ \ Physics $ }
\centerline{$ \it ^{ b}\  Moscow \ \ Physical \ \ Technical \ \ Institute $}
 \centerline{$ \it ^{c} \  Moscow \ \ State \ \ University $}

\bigskip

\begin{abstract}
The lagrangian of the  N=1, D=10 dual supergravity interacting
with the Yang-Mills matter multiplet is constructed starting
immediately from the equations of motion obtained from the
Bianchi Identities in the  superspace approach. The difference
is established in comparison with the Gates-Nishino lagrangian
at the forth order level in fermionic fields.
\end{abstract}

\section{Introduction}

The dual version of ten-dimensional (D=10) simple (N=1)
super\-gravity (DU\-AL SUGRA) was discussed in \cite{C}, \cite{GN1}.
(The complete lagrangian was written  in \cite{GN1}).  Some
arguments were presented in
\cite{GN2}, that this theory may be considered  as a low-energy
limit of the geterotic superstring. In the same paper  special
nonlocal interactions at the superstring level were constructed
to reproduce  the DUAL SUGRA in the string theory. It
was shown in \cite{D}, that DUAL SUGRA may be introduced naturally
 in the framework of a five-brane
theory, which is apparently related by duality to the theory of
superstring. The connection of the DUAL SUGRA with the
superstring (and/or five-brane) approach is important for many
reasons. In particular, the special nonlinear superstring
corrections are necessary for compensation of anomalies.

 Unfortunately, the five-brane theory has not yet studied
properly. But in the superstring approach one encounters
complicated nonlocal interactions in the derivation of  DUAL
SUGRA.  So,  it is not clear how to obtain the DUAL SUGRA
anomaly-free lagrangian  from the five-brane and/or superstring
approach even in the tree approximation.

   Nevertheless  it is possible to construct anomaly-free DUAL
SUGRA immediately  using  symmetry
considerations only .  One must use in addition  some knowlege
on the structure of the Green-Schwarz \cite{GS} anomaly
compensating terms which arise in the proccess of dual
transformation from the USUAL SUGRA theory. This approach was
accepted in \cite{T1} and it was shown finally in \cite{T2} that
equations of motion (e.m.'s) of the resulting theory are much
more simple than in the USUAL  SUGRA (where e.m.'s has no closed
form and can be presented only as an infinite series in terms of
3-form graviphoton superfield, see
\cite{MANY} and other references therein). Unfortunately
the discussed approach is based on superspace methods and
corresponds specifically to the mass-shell description, i.e. to
the e.m.'s level.

 The construction of the lagrangian is a separate problem. In
general it is not clear a priori that this problem may be solved
in the  anomaly-free case,  where a lot of nonlinear terms
(tree-level superstring corrections) are present. The absence of
the explicit e.m.'s  makes it difficult  to study  this problem
in the USUAL SUGRA case. Another situation is in the DUAL SUGRA,
where one may hope to construct the lagrangian explicitely. That
is our final purpose in the process of studying of the DUAL
SUGRA.

   To solve this problem one must study first the simplest
(anomaly-full)  case of  DUAL SUGRA (without nonlinear
superstring corrections). The lagrangian of such a theory has
been  derived from the USUAL SUGRA
lagrangian
\cite{WN},\cite{CM}
by an explicit dual transformation, see \cite{GN1}.  (This
method can not be generalized to the anomaly-free case).  Our
purpose here is to derive this lagrangian immediately from the
e.m.'s in the superspace approach, because we believe that
namely this approach can be generalized to the more general
anomaly-free case. The main idea of this derivation was
formulated in the short paper by one of us \cite{Z}, where the pure
gravity sector  was considered.  In the present
paper the straightforward generalization to the case of gravity
interacting with matter is considered.  We found out in the
process of this study that some fermionic forth
order  terms  are given incorrectly in \cite{GN1}.

 In Sec. 2 the supercovariant e.m.'s for matter fields are
presented, in Sec. 3 the lagrangian for matter sector is
derived from these equations.  In this section the lagrangian
for gravity sector is also presented,  which is taken from
\cite{Z}.

In Sec. 4 the e.m.'s for gravity sector (including matter
contribution) is derived from the superspace aproach and the
derivation of these equations from the
lagrangian is also discussed. That provides the complete check of the
procedure.  In Sec. 5 the supersymmetry
transformations are discussed.

We use the special field  parametrization  \cite{T2} (which is
closely related to the parametrization in \cite{N}), which
greatly simplify calculations.  In Sec. 6  the
super-Weil transformation is discussed, connecting our fields
 to the set of
fields from \cite{GN1} and \cite{CM}.

 In Sec. 7 we discuss the scale-invariance, which will be
important in analysing the general structure of the lagrangian
with superstring/fivebrane corrections  at the next stage of our
work. This invariance is also helpful in establishing of general
structure of Bianchi Identities (BI's) in the superspace
approach.

In the Appendix the supercovariant form of the e.m.'s
 is presented  together with some important
constraints for superfields.

Finally the additional comment is needed.  The most of the
studies in the  superspace approach stop at the level of
equations of motion. There are reasons for this.  These e.m.'s
 are written usually  in terms of torsion-full spin-connection and
fields are defined in the superspace covariant notations.  So,
there is a long way from this level of description to the
lagrangian level. But it is a technical problem. What is more important
that the presence of additional constraints in the superspace approach
 complicates the transition from the e.m.'s to the
lagrangian.  With some parametrizations  used in the
literature (i.e. with some choice of the constraints) it is even
not clear, that the lagrangian level  may be consistently
realized.  Our purpose in this paper is to go along this way
from the beginning to the end.

  Our notations and conventions in general are the same as in
\cite{Z}.  (Some small differences will be  noted below). These
notations correspond to \cite{T2} up to the sign of curvature
and spin-connection.  The complete description of the notations
and their connections to that from other papers see in
\cite{T1}.

\section{Supercovariant Equations of Motion for Matter Fields }

The derivation of the matter fields e.m.'s is the   standard
procedure in the superspace approach. (For example see
\cite{BBLPT}, \cite{ADR} and references therein).  We are
presenting here some basic formulas only  to define our
notations.

The starting point is the BI for the Yang-Mills field-strength
superfield ${\cal F}_{AB}$:

$$ D_{[A}{\cal F}_{BC)} + {T_{[AB}}^Q\, {\cal F}_{QC)}  \equiv 0,
                   \eqno(2.1) $$
where supercovariant derivatives ${D_A}$ obey the commutative
(anticommutative) relations:

$$ (D_A\, D_B - (-1)^{ab}D_B\, D_A)\, V_C = $$
$$ = -\, {T_{AB}}^Q\, D_Q \,V_C -
{{\cal R}_{ABC}}^D\, V_D -({\cal F}_{AB}\,V_C -
(-1)^{c(a+b)}V_C\,{\cal F}_{AB}),
    \eqno(2.2)  $$
where the supertorsion $ T_{AB} $  is defined as in \cite{T2},
but the supercurvature ${\cal R}_{ABCD} $ differs in sign in comparison
with \cite{T2}; $ {\cal
F}_{AB} $ is in the algebga of the Yang-Mills internal symmetry
group $G$: ${\cal F}_{AB}  \equiv {\cal F}^J_{AB}X^J $, where
${(X^J)_i}^j $ are anti-hermitian matrices - generators of $G$.
Our definition corresponds to the space-time
components of the Yang-Mills field-strength in the form: $F_{mn}
= \partial_m A_n - \partial_n A_m - [A_m, A_n], $  where $A_m$
is the space-time components of vector-potential. That corresponds to
 $F=dA+A^2 $
for the field-strength  2-form. The same definition is accepted for
the curvature 2-form: $dR = dw + w^2, $ where $w$ is the
spin-connection 1-form.

We use the same set of constraints as in
\cite{T2}, so we get the same solution for the torsion and
curvature BI's.

    To find the mass-shell solution of  BI (2.1) it is
necessary to impose the additional constraint:

  $${\cal F}_{\alpha\beta} = 0                  \eqno(2.3)      $$
Then, using  the the standard
procedure (cf. \cite{ADR} and other references therein) one can
derive the relations which follow from (2.1) and (2.3):

$${\cal F}_{a \alpha}\equiv
{(\Gamma_a)}_{\alpha\beta}\,\lambda^\beta,       \eqno(2.4) $$
$$ {D_\alpha}\,\lambda^\gamma={1\over4}\,{{\cal F}_{ab}}\,
{{(\Gamma^{ab})}_\alpha}^\gamma                    \eqno(2.5) $$
$$D_\gamma\,{\cal F}^{ab}=2{(\Gamma^{[a})}_{\gamma\beta}\,
D^{b]}\,{\lambda}^\beta-
T^{abc}\,{{(\Gamma_c)}_{\gamma\beta}}\,{\lambda}^{\beta}
-{1\over36}\,
{(\hat {T}\Gamma^{ab})}_{\gamma\beta}\,{\lambda}^{\beta}     \eqno(2.6)$$
where $\lambda^\alpha $ (which is a 16 IR of $O(1.9)$ ) must be
interpreted as the gluino superfield.

Now, applying  spinorial derivatives to  eq. (2.5), then
taking the symmetrical part   in spinorial  indices  of the
resulting expression  and  using (2.2), (2.6),  one gets the
following equations of motion:

$$ \Gamma^a\,D_a\,\lambda + {1\over12}\, T_{abc}\Gamma^{abc}\, \lambda
 = 0, \eqno(2.7) $$

$$ D_a\,{\cal F}^{ab}+
T_{ab}\Gamma^b\,\lambda+2\,\lambda\,\Gamma^b\,\lambda=0.  \eqno(2.8)$$
We do not write spinorial indices explicitely in the cases, where their
position may be reconstructed unambiguously.

Taking the zero superfield-component of these equations,  one gets
immediately the e.m.'s for gluon and gluino fields in supercovariant
notations. (In the following
we make no difference for the relations between superfields and their
zero components because in  most of the cases it can not
produce misunderstanding).

\section{Lagrangian for Physical Matter Fields }

The simple equations obtained in  Sec. 2  are not suitable
for construction of the lagrangian,  because they are written in
terms of superspace-covariant tangent-space components. (We are
not able to write a lagrangian in terms of these components).
As usual, to return to the space-time components one
must use the special gauge
for the superspace veilbein
$E_M^A $  (cf.  \cite{BW}):

$$ {E_M}^A \vert =
\left(
\begin{array}{ll}

 {e_m}^a & \psi_m^\alpha \\
 0     & \delta_\mu^\alpha

\end{array} \right)\, ,      \eqno(3.1)    $$
where $\psi_m^\alpha $ is the gravitino field.

The supercovariant derivative $D_a \equiv {E_a}^M\,D_M $ is equal to:

$$D_a = e_a^m\,D_m -\psi_a^\beta\,D_\beta \, ,   \eqno(3.2)$$
where $\psi_a = e_a^m\, \psi_m $ and
the space-time component of the covariant
derivative is:

$$ D_m \lambda = \partial\, \lambda - \omega_m \, \lambda -
[A_m, \lambda],                                \eqno(3.2')       $$
where ${(\omega_m )^\beta}_\gamma \equiv {1\over 4}
{\omega_m}^{ab}{{\Gamma_{ab}}^\beta}_\gamma   $ is the
spin-connection which is in the algebra of $O(1.9)$.

Now we introduce the usual tangent-space components of physical
fields  instead of supercovariant quantities used in the
superspace approach of refs. \cite{T2} and others.  Namely:

$$F_{ab} \equiv e_a^m\,e_b^n\,F_{mn}, \ \ \ \
\omega_{cab} \equiv e_c^m\, \omega_{mab}\, ,   \eqno(3.3)                $$

$$M_{abc}={1\over7!}\,{\varepsilon_{abc}}^{a_1 \ldots a_7}\,({e_{a_1}}^{m_1}
\,\ldots {e_{a_7}}^{m_7}\, N_{m_1\ldots m_7})\, ,  \eqno(3.4) $$
where $ N_{m_1\ldots m_7}= 7\, \partial_{[m_1}\,M_{m_2 \ldots m_7]}$, and
 $M_{m_1\ldots m_6}$ is the 6-form graviphoton potential of DUAL
SUGRA.

It is  possible by the standard way, using the definition of
supertorsion $ {T_{MN}}^A = D_M\,E_N^A - (-1)^{mn}\,D_N\,E_M^A\, ,$
 to find the relation between the torsion-full
spin-connection in the eq.(3.3) and the usual spin-connection
$\omega_{cab}^{(0)}$ defined in terms of derivatives of $e_m^a$:

$$ \omega_{cab}= \omega_{cab}^{(0)}(e)   +{1\over2}\,T_{cab}+C_{cab}\, ,
\eqno(3.5)$$ where:
  $$ C_{cab}= \psi_a\,\Gamma_c\,\psi_b - {3\over 2}
\psi_{[a}\,\Gamma_c\,\psi_{b]}    \eqno(3.5')$$

We need the special notation $  \nabla_m   $ for the covariant
derivative with the spin-connection $\omega_m^{(0)}$
($ \nabla_{[m}e_{n]}^a =0$).  We define also $\nabla_a \equiv e_a^m
 \nabla_m $.

 Now it is the straightforward procedure to connect the physical fields
introduced before with ${\cal F}_{ab}$ and  other supercovariant fields
from \cite{T2}:

$${\cal F}^{ab}=F^{ab}+2\,\psi^{[a}\,\Gamma^{b]}\,\lambda   \eqno(3.6)$$

$$T_{ab} = 2\,\nabla_{[a}\,\psi_{b]}+
{1\over2}\,(\Gamma^{cd})\,\psi_{[a}\,C_{b]cd}  \eqno(3.7)$$

$$T_{abc}= N_{abc} =
M_{abc}-{1\over2}\,\psi_f\,{{\Gamma^f}_{abc}}^d\,\psi_d
\eqno(3.8)$$

Substituting  obtained relations in the eqs.(2.7),(2.6) and
taking into account the  eq. (3.2),   we get equations of motion for physical
matter  fields in  the final form.
For gluino:
$$ {\hat \nabla}\,\lambda
-{1\over24}\,\hat{M}\,\lambda+
{1\over48}\,(\psi^f\,\Gamma_{fabcd}\,\psi^d)\,\Gamma^{abc}\,\lambda+
{1\over4}\,\Gamma^a\,\hat{F}\,\psi_a +$$
$$+{1\over8}\,(\psi_a\,\Gamma_b\,\psi_c)\,\Gamma^{abc}\,\lambda
-{1\over2}\,(\psi_a\,\Gamma^a\,\lambda)\,\Gamma^b\,\psi_b+
{1\over2}\,(\psi_m\,\Gamma_a\,\lambda)\,\Gamma^a\,\psi^m=0,  \eqno(3.9)$$
For gluon:
$$ \nabla_b\,(F^{ba}-\lambda\,\Gamma^c\,\Gamma^{ba}\,\psi_c)+
2\,\lambda\,\Gamma^a\,\lambda+{1\over2}\,M^{abc}\,F_{bc}=0\, ,  \eqno(3.10)$$
where
$$ {\hat \nabla}=\Gamma^a\,\nabla_a,\ \ \
\hat{F}=F_{ab}\,\Gamma^{ab}, \ \ \ {\hat M} = M_{abc}\Gamma^{abc}
\eqno(3.10') $$

The matter-field lagrangian-density is reconstructed
 immediately from  eqs.(3.9),
(3.10):
$$ {\cal L}_{YM}={1\over g^2}\, Tr[-{1\over4}\,F_{ba}\,F^{ba}+
{1\over{8 \cdot 6!}}\,{\varepsilon}^{a_1 \ldots a_{10}}\,M_{a_1 \ldots a_6}
\,F_{a_7a_8}\,
F_{a_9a_{10}} +$$
$$+\lambda\,{\hat \nabla}\,\lambda
-{1\over24}\,\lambda\,\hat{M}\,\lambda
+{1\over2}\,\lambda\,\Gamma^a\,\hat{F}\,\psi_a$$
$$+ (\lambda\, \Gamma_b\, \psi_a)\,(\lambda\, \Gamma^a\, \psi_b)
-{1\over2}\,(\lambda\,\Gamma^b\,\psi_b)^2-
{1\over2}\,(\lambda\,\Gamma_a\,\psi_b)^2],    \eqno(3.11)$$
where $Tr$   is calculated in
the adjoint representation of the group G
 ($ Tr \,AB = G_{JK}\,A^J\,B^K $ ,
where $G_{JK}$ is a Killing tensor), $g$ is a coupling constant.

 The total lagrangian is equal to (in all the cases here and in the following
the term "lagrangian" is used for the lagrangian density):

 $$ {\cal L}_{tot} = {\cal L}_{YM} + {\cal L}_{GRAV}\, ,  \eqno(3.12) $$
 where the first term in the r.h.s. of (3.12) is the lagrangian
(3.11), but the second term is the lagrangian for the pure
gravity supermultiplet, obtained in \cite{Z}:

$$ {\cal L}_{GRAV} =
\phi\,({\cal R}-{1\over3}\,T^2)\, \vert +2\chi\,\Gamma^{ab}T_{ab}\, \vert \, ,
 \eqno(3.13) $$
where $\phi$ and $\chi_\alpha $ are dilaton and dilatino fields
($\alpha$ is spinorial index),  ${\cal R}$ is the supercovariant
scalar curvature, $T_{ab}^\alpha$ and $T_{abc}$ are
supercovariant torsion components, $T^2 \equiv T_{abc}\,T^{abc}, \
$ gravitational coupling constant is put equal to one.

This form of the lagrangian follows from the linearity of the
superspace  e.m.'s
 in terms of dilaton and dilatino fields (see
Appendix).

It is not a direct procedure to obtain from (3.13) the
lagrangian in terms of physical fields. (The presence of
constraints at  the on-shell level makes it difficult to relate the
off-shell curvature in (3.13) with the corresponding on-shell
quantity, defined in \cite{T2}).  This problem was solved  in
\cite{Z} and the resulting lagrangian takes the form:

$$ {\cal L}_{GRAV} =
 \phi\,R  - 2\,\phi\,\psi_a\Gamma^{abc}\psi_{c;\,b} -
 2\,\phi_{;\,a}\psi^a\Gamma_b\psi^b +
4\, \psi_a\Gamma^{ab}\chi_{;\,b} +                                           $$
$$- {1\over 12}\,\phi\,M_{abc}^2 + {1\over 12}\,\phi\,\psi_a\Gamma^{[a}
{\hat M}\Gamma^{b]}\psi_b - {1\over 2}\,\chi\,\Gamma^{ab}\psi^c M_{abc} -   $$
$$ - {1\over 48}\,\phi\,{(\psi^d\Gamma_{dabcf}\psi^f)}^2 +
{1\over 4}\,\phi\,{(\psi_a\Gamma_b\psi_c)}^2 +
{1\over 2}\,\phi\,(\psi^a\Gamma^b\psi^c)(\psi_a\Gamma_c\psi_b) -           $$
$$ - \phi\,{(\psi_a\Gamma^b\psi_b)}^2 +
(\chi\Gamma_{ab}\psi_c)(\psi^a\Gamma^c\psi^b) -
2\,(\chi\Gamma_a\Gamma_b\psi^b)(\psi^a\Gamma_c\psi^c)    \eqno(3.14) $$

 Some  specific property of  the
graviphoton $M_{m_1\ldots m_6}$-field equation of motion (namely the fact
that this equation has the form of derivative, because only the
field-strength
appears in the lagrangian) is
necessary to derive (3.14) from (3.13)
\cite{Z}, i.e.  only partial information on the superspace
e.m.'s
 is sufficient for the construction of
the complete lagrangian.  Afterwards,  having (3.11), (3.12) and
(3.14), one may write  explicitely all the other equations of motion.

 To check the consistency of the approach  we take the following
procedure  in the next Section . We derive the explicit form of
the superspace e.m.'s  and then compare the result with the
e.m.'s which follow from the lagrangian. We shall find that the
lagrangian e.m.'s  are, in general, complicated linear
combinations of the superspace e.m.'s.  Nevertheless there is
the complete corresponence between them (see below).

The additional comment is necessary. The lagrangian (3.11) does
not contain terms of order of $ \lambda^4 $. (The choice of
 variables corresponding to the canonical kinetic terms in the
lagrangian does
not change this conclusion, see Sec. 6). That  contradicts to the
result of \cite{GN1}. (We find also the discrepancy with
\cite{GN1} in some other forth order  terms in fermionic
fields).

\section{Gravity Multiplet Equations of Motion}.

The complete set of the  superspace e.m.'s, as derived in
\cite{T2}, is presented in the Appendix. They have the universal form
for
DUAL and USUAL SUGRA.  These equations contain the $A_{abc}$
-field, which, at the first sight, is not fixed completely
in the DUAL SUGRA. Nevertheless, we believe that equations (A.11),
 (A.12) fix the $A_{abc}$-field unambiguously up to some numerical
multiplicative factor. The simplest way to find the solution of
these equations ( it is the unique
solution consistent with BI's), is to consider  BI's for the graviphoton field
$H_{ABC}$  in the USUAL SUGRA. Such a procedure was used in
\cite{T2} to find the contribution of superstring corrections.
Now we use it to find the contribution of matter
degrees of freedom to the gravity multiplet e.m.'s.

In the presence of matter fields the $H$-superfield BI's
take the form:

$$ D_{[A}H_{BCD)} +{3\over2}\,{T_{[AB}}^Q\, H_{QCD)} =$$
$$ = - {3\over 4}\, c_Y\,
Tr[{\cal F}_{[AB}\,{\cal F}_{CD)}] -
{3\over 2}\, c_L \, {R_{[AB}}^{ef}{R_{CD)}}_{ef}     \eqno(4.1) $$
Note, that $ c_L = -\gamma \neq 0,\ \ c_Y =0 $ in \cite{T2}.
Now we are considering the case $c_L = 0, \ \ c_Y \neq 0$.

The  factor $c_Y$ is fixed in the USUAL SUGRA if the $H$-field
normalization   is fixed by the choice of
constraints (4.2) (see below) but the ${\cal F}$-field normalization
 is fixed by the choice of kinetic terms in the
lagrangian.  The value of $c_Y$ also follows
 in the framework of the DUAL SUGRA (see
below, eq. (4.11)).  Due to the fact, that $H_{abc}$ and
$M_{abc}$ are connected   with the same quantity,
 the torsion-component $T_{abc}$, one can easily relate the
normalization of $H$ and $M$-fields. So one can easily establish
the self-consistency of the $c_Y$-definition by two different
procedures.

The  constraints for the $H$-superfield may be
self-consistently defined in terms of dilaton  $\phi$ and
dilatino field  $ \chi=D\phi $ in the form:

$$ H_{\alpha\beta a}=\phi\,(\Gamma_a)_{\alpha\beta}\; ,  \eqno(4.2) $$

Using (4.2), one can find the solution of (4.1) which is
consistent with the solution of torsion BI's in \cite{T2}.  The
result is:

$$ A_{abc}= - {c_Y\over 96}\,Tr[\lambda\,\Gamma_{abc}\,\lambda], \eqno(4.4)$$
$$ H_{\alpha bc}=-(\Gamma_{bc}\,\chi)_\alpha ,         $$
$$H_{abc}=
-\phi\,T_{abc}-{c_Y\over 4}\,Tr[\lambda\,\Gamma_{abc}\,\lambda] \eqno(4.5)
$$

One may check that (4.4) is also the explicit solution of the
 $A$-field equations (A.11), (A.12) (see  Appendix).
Now one may forget the USUAL SUGRA, considering the described
procedure as the helpful auxiliary method to find the
$A_{abc}$-field explicitely, - not more.

Then it is a straightforward procedure to write the superspace
equations of motion (A.6)-(A.10)  in terms of physical fields
entering in the lagrangian.  The calculations and the results are
rather cumbersome. We present here only the terms
which come from the matter-fields contribution via the
$A_{abc}$-tensor in the e.m.'s  (A.6)-(A.10).  (The pure gravity
contribution was discussed in \cite{Z}).  We get the equations:

for the gravitino:

$$ Q_a \equiv \phi\,T_{ab}\Gamma^b - \nabla_a \chi    +  \ldots +$$
$$+{c_Y \over
96}\,Tr[(\lambda\,\Gamma_{bcd}\,\lambda)\,\Gamma^{bcd}\,\psi_a
+4\,F_{cd}\,{\Gamma_a}^{cd}\,\lambda+8\,(\psi_c\,\Gamma_d\,\lambda)\,
{\Gamma_a}^{cd}\,\lambda -$$
$$-40\,F_{ac}\,\Gamma^c\,\lambda
-80\,(\psi_a\,\Gamma_c\,\lambda)\,\Gamma^c\,\lambda] = 0,  \eqno(4.6) $$
(here and in the following the   notation $\ldots$ is used for the
pure gravity  contribution, nonlinear in fields),

for the dilatino:

$$Q \equiv  {\hat \nabla}\,\chi + \ldots -$$
$$-{c_Y\over 96}\,Tr[(\lambda\,\Gamma_{bcd}\,\lambda)\,\Gamma^a\,
\Gamma^{bcd}\,\psi_a
-8\,\hat{F}\,\lambda-16\,(\psi_a\,\Gamma_b\,\lambda)\,\Gamma^{ab}\,\lambda]
=0,   \eqno(4.7)$$

for the dilaton:

$$ S \equiv \nabla_a \nabla^a\,\phi + \ldots  + $$
$$+ {c_Y\over 96}\,Tr[(\lambda\,
\Gamma_{abc}\,\lambda)\,(\psi_f\,\Gamma^{abc}\,\psi^f)-
8\,(F^{ab}\,F_{ab})-$$
$$- 32\,(\psi_a\,\Gamma_b\,\lambda)\,F^{ab}
-32\,(\psi_{[a}\,\Gamma_{b]}\,\lambda)^2
+{4\over3}\,(\lambda\,\hat{T}\,\lambda)] = 0,    \eqno(4.8)$$

 for the graviton:

$$ S_{ab} \equiv \phi\,{\cal R}_{ab} +
 \nabla_{(a} \nabla_{b)}\phi + \ldots +$$
$$+ {c_Y\over 96}\, Tr[(\lambda\,
\Gamma_{cde}\,\lambda)\,(\psi_a\,\Gamma^{cde}\,\psi_b)
+4\,\eta_{ab}\,(F^{cd}\,F_{cd})
+
16\,\eta_{ab}\,(\psi_c\,\Gamma_d\,\lambda)\,F^{cd} +$$
$$+16\,\eta_{ab}\,
(\psi_{[c}\,\Gamma_{d]}\,
\lambda)^2-48\,(\nabla_{(a}\,\lambda)\,\Gamma_{b)}\,\lambda
+6\,(\lambda\,{\Gamma^{cd}}_{(b}\,\lambda)\,T_{a)cd}-$$
$$-12\,(\lambda\,{\Gamma^{cd}}_{(b}\,\lambda)\,C_{a)cd}
+12\,\psi_{(a}\,\hat{F}\,\Gamma_{b)}\,\lambda
+24\,(\psi_{(a}\,{\Gamma^{ij}}_{b)}\,\lambda)\,(\psi_i\,\Gamma_j\,\lambda) -$$
$$-{2\over3}\,\eta_{ab}\,(\lambda\,\hat{T}\,\lambda)
+48\,F_{ac}\,{F^c}_b+96\,(\psi_{[a}\,\Gamma_{c]}\,\lambda)\,{F^c}_b
+96\,(\psi_{[b}\,\Gamma_{c]}\,\lambda)\,{F^c}_a +$$
$$+120\,(\psi_i\,\Gamma_{(a}\,\lambda)\,(\psi_{b)}\,\Gamma^i\,\lambda)
-72\,(\psi_a\,\Gamma^j\,\lambda)\,(\psi_b\,\Gamma_j\,\lambda)-$$
$$-48\,(\psi_c\,\Gamma_a\,\lambda)\,
(\psi^c\,\Gamma_b\,\lambda)] = 0, \eqno(4.9) $$

 for the graviphoton:

$$ S_{abcd} \equiv \nabla_{[a}\,(\phi\,M_{bcd]})+\ldots +$$
$$+{c_Y\over 8}\,Tr[
-6\,F_{[ab}\,F_{cd]}-24\,(\psi_{[a}\,\Gamma_b\,\lambda)\,
(\psi_c\,\Gamma_{d]}\,\lambda)-$$
$$-24\,(\psi_{[a}\,\Gamma_b\,\lambda)\,F_{cd]}
+4\,\nabla_{[a}\,\lambda\,\Gamma_{bcd]}\,\lambda
- $$
$$ -6\, \lambda\, {\Gamma^j}_{[ab}\, \lambda\, C_{cd]j}
-\psi_{[a}\,\hat{F}\,\Gamma_{bcd]}\,\lambda
-2\,(\psi_{[a}\,\Gamma^{ij}\,\Gamma_{bcd]}\,\lambda)\,(\psi_i\,\Gamma_j\,
\lambda)]=0,  \eqno(4.10) $$

These equations may be derived independently by  variation of
the lagrangian (3.12).  (matter-field contribution to them follows
from  (3.11)).  The result of this variation is consistent with
 eq.'s (4.6)-(4.10) if:

$$ c_Y = { 1\over g^2}.         \eqno(4.11) $$

But there is no direct
correspondence between (4.6)-(4.10) and the equations obtained
by variation of the lagrangian.

  Note from the beginning
that the dilaton eq.(4.8) immediately  folows from (4.9) due to
the constraint (A.4c) ( one must multiply (4.9) by
$\eta^{ab}$ to get (4.8)); the dilatino eq. (4.7) follows from
(4.6) due to the constraint (A.4b),  \cite{T2} (one must
multiply (4.6) by $\Gamma^a$ to get (4.7)).

The variation of (3.12) with respect to the
 gravitino field $\psi_m$ produces the
equation:

$$ Q_a + \Gamma_a \, Q =0.          \eqno(4.12)   $$

The variation of (3.12) with respect to the graviphoton field
$M_{m_1\ldots m_6}$ produces the equation:

$$ S_{abcd} + 3\, \psi_{[a}\, \Gamma_{bc}\, Q_{d]} = 0. \eqno(4.13) $$

The variation of (3.12) with respect to the graviton field
$e_m^a$ produces the equation

$$ S_{ab} + \eta_{ab}\, ({1\over 2}B-S)-2\psi_{(a}Q_{b)}-
{1\over 2}\psi^c \Gamma_{ab}\,Q_c - \psi^c \Gamma_{c(a}Q_{b)} -$$
$$ -{1\over 2}(\psi^c \Gamma_{cab} + 2\psi_a\Gamma_b)Q
- \eta_{ab}\,  \psi_c \Gamma^c Q
 -{1\over2}\, \eta_{ab}\, Tr(\lambda \Lambda)-
{1\over4}\, Tr(\lambda \Gamma_{ab} \Lambda) = 0.
                                          \eqno(4.14) $$
where $ B\equiv  -\phi ({\cal R} - {1\over 3}\, T^2), \ $ but
  $ \Lambda \equiv ({\hat \nabla} \lambda + \ldots) =0 $ is
the l.h.s. of the gluino equation (3.9),
 $Q,\  Q_a,\  S, \ S_{ab},\  S_{abcd} $ are defined by (4.6)-(4.10).

 The direct variation of (3.12) with respect to the dilaton $\phi$
and the dilatino $\chi$-fields produces the constraints (A.4b),
(A.4c) as it follows from
(3.13). (Note that $\phi$ and $\chi$ does not
enter into the matter part of the lagrangian (3.11)). So $B=0$
in (4.14).

Calculating $\Gamma_a $ projection from (4.12) one immediately
obtains $Q=0$, and then $Q_a=0$. So, $ S_{abcd}=0 $ as it follows
from (4.13).  Contracting $a,b$ indices in (4.14) one obtains
$S=0$, and then $S_{ab}=0$.
  So, all the equations (4.6)-(4.10)
follow from (4.12)-(4.14).

This discussion demonstrates the complicated inter-connection
between the lagrangian and the superspace  e.m.'s.
  It is the price one must pay for the
simplicity of the superspace mass-shell formulation.

The  consideration of pure gravity sector (terms $\ldots$ in
(4.6)-(4.10) ) leads to the same equations (4.12)-(4.14).
(This calculation was done by one of us (K.N.Z) and provide a
check of the procedure).

It is important that the same combinations (4.12)-(4.14) must
follow from the variation of the lagrangian if the contribution
of superstring corrections is taken into account according to
\cite{T2} (if a lagrangian exists in this case) because, as it has been
just shown, only consideration of matter-gravity interaction terms is
sufficient for the derivation of (4.12)-(4.14).  This
observation must help the construction of the lagrangian from
the superspace e.m.'s in the presence of superstring corrections.

\section{Supersymmetry Transformations}

The supersymmetry transformations for any physical field follows
 immediately  from the super-gauge transformation for the
corresponding superfield (cf. \cite{BW}). Out definitions are the following.

 The super-gauge transformation is:

 $$ \delta_Q(\epsilon) = \delta_{GCT}(\xi^N) + \delta_L (L_{ab}) +
\delta_G,  \eqno(5.1) $$
 where $\delta_{GCT} $ is a special superspace general  coordinate
transformation:

$$ \delta_{GCT}(\xi^N) V_M =
- \xi^N \partial_N V_M - \partial_M \xi^N V_N \, ,
\eqno(5.2) $$
where  $\xi^N = (\epsilon^\nu, 0) $ is a parameter, $V_M$ is
any field with a world-index in the superspace;

 $\delta_L $ is a Lorentz-transformation:

$$ \delta_L(L_{ab}) F = -(L_{ab}\,{\hat M}^{ab}) \, F,         \eqno(5.3) $$
where $L_{ab}$ are parameters, ${\hat M}^{cd} $ are
Lorentz-group generators, $F$ is any field with
a tangent-space
index in
 the super\-space.
   (Our defi\-ni\-tions are:
    ${\hat M}^{cd}\,
\lambda^\alpha = {1\over4}
  {(\Gamma^{cd})^\alpha}_\beta
\,\lambda^\beta $ and ${\hat M}^{cd}\, X_a  = \delta_{[ab]}^{\, cd}\,
X^b $ for fields
 with spinorial and vector indices);

$ \delta_G$ is a  gauge transformation:

$$ \delta_G(\Omega) \, A_m = - [\Omega, \, A_m]- \partial_m\, \Omega\, ,
 \eqno(5.4')  $$
$$ \delta_G(f)M_{n_1\ldots m_6} = -6\, \partial_{[n_1}f_{n_2\ldots n_6}
  \eqno(5.4'') $$
 where $\Omega$ and $f_{n_1\ldots n_5}$ are  gauge transformation parameters.

  One can easily find  (using the
standard procedure \cite{BW})    all the
parameters in (5.3), (5.4)  from the condition, that only
  the superveibein   transformation contains the
derivative of $\epsilon$.  We find:

 $$ L_{AB} =- \epsilon^\nu \omega_{\nu AB}, \ \ \ \ \Omega =
-\epsilon^\nu A_\nu, \ \ \ \
 f_{n_1\ldots n_5}= - \epsilon^\nu M_{\nu n_1\ldots n_5}   $$
Then we get for the super-veilbein:

  $$ \delta_Q(\epsilon) E_M^A =- D_M\xi^A - \epsilon^\alpha T_{\alpha M}^A\,  ,
  \eqno(5.5)$$
where $ \xi^A = (\epsilon^\alpha, 0) ;\ $
  for any gauge-covariant field:
  $$ \delta_Q(\epsilon) \, X = -\epsilon^\alpha D_\alpha  X    \eqno(5.6)   $$
  and for field-potentials:
  $$ \delta_Q(\epsilon) \, A_m = -\epsilon^\alpha F_{\alpha m}   \eqno(5.7') $$
  $$ \delta_Q(\epsilon) \, M_{n_1 \ldots n_6} =
  -\epsilon^\alpha N_{\alpha n_1\ldots n_6}       \eqno(5.7'')  $$

 We write here the final form of the supersymmetry transformation in
terms of  superfields. The result for zero components (physical fields)
 follows
immediately with the help of relations from sec.'s 2-4.

For matter multiplet:

 $$ \delta_Q(\epsilon) \, \lambda =
  {1\over4}\, {\cal F}_{ab} \, \Gamma^{ab}\, \epsilon   \eqno(5.8a)  $$
 $$ \delta_Q(\epsilon) \, A_m = -\lambda \, \Gamma_m \, \epsilon,  \eqno(5.8b)
$$
 (where as usual $\Gamma_m \equiv e_m^a \,  \Gamma_a \ $);
for  gravity multiplet :

$$\delta_Q(\epsilon) {e_m}^a =- \psi_m\Gamma^a\epsilon\, , \eqno(5.9a)    $$
$$ \delta_Q(\epsilon) \psi_m =- D_m \epsilon - {1\over 72}\,
 \Gamma_m{\hat T}\,\epsilon\, ,       \eqno(5.9b)                   $$
$$ \delta_Q(\epsilon) \phi =  \chi\,\epsilon\, ,  \eqno(5.9c)               $$
$$ \delta_Q(\epsilon) \chi =  {1\over 2}\, D_a  \phi \,\Gamma^a\epsilon -
({1\over 36}\,\phi{\hat T} -
{\hat A} )\, \epsilon\, , \eqno(5.9d)                   $$
$$ \delta_Q(\epsilon) M_{m_1 \ldots m_6} =  6\,\psi_{[m_1}
\Gamma_{m_2 \ldots m_6]}\,\epsilon\,                 \eqno(5.9e) $$
where ${\cal F}^{ab}$
is defined in (3.6); ${\hat T} = T_{abc}\, \Gamma^{abc}$,
 (the same for ${\hat A}$),
$T_{abc}$ is defined in (3.8), $A_{abc}$ is defined in (4.4).

 The additional terms
should be included in the $A_{abc}$-field if superstring corrections are
present, cf. \cite{T2}.  It is the advantage of our
parametrization, that matter degrees of freedom (as well as
superstring corrections) "penetrate" the gravity multiplet
supersymmetry transformations only due to the
$A_{abc}$-contribution as in (5.11).

The supersymmetry algebra for physical fields
is closed up to  equations of motion
and  gauge transformations. Namely:

$$ [\delta_Q(\epsilon_2),\, \delta_Q(\epsilon_1)]\,X =
( \delta_{GCT}(\xi^m)+  \delta_Q(\epsilon') + $$
$$  + \delta_L(L_{ab})
+ {\delta_G}(\Omega_{YM}) + {\delta_G}(f_{n_1\ldots n_5})
  )\, X +
 (\mbox{e.m.'s}),           \eqno(5.10) $$
where $X$ is any field from gravity or  matter multiplet.

The  transformation parameters  in (5.10) are:

$$ \xi^m = \epsilon_1\, \Gamma^m \, \epsilon_2\, .       $$
$$ \Omega_{YM} = -  \xi^m \, A_m\, .     $$
$$ \Omega_{m_1,\ldots, m_5} = -\xi^n\,M_{m_1,\ldots, m_5,n}  $$
$$ L_{ab} = - \xi^n \, \omega_{nab} + {5\over 12}\, \xi^c\, T_{abc} +
{1\over 36}\, \epsilon_1\, {\Gamma_{ab}}^{cde}\, \epsilon_2\, T_{cde}.
  $$
$$ \epsilon' = \xi^n\, \psi_n                     $$
 Eq. (5.10) takes place for any $A_{abc}$-field (not
specifically for that, defined by eq.(4.4)).  Only the
representation (A.13) for the $A_{abc}$-superfield spinorial
derivative is necessary  for the derivation of (5.10) .

\section{Super-Weil transformations}

To find the natural variables, where the lagrangian  is  more
complicated, but all the kinetic terms have a canonical
structure, the corresponding nonlinear transformation of the
fields must be established.  The most important part of this
transformation was found in  \cite{Z}  by a direct
study of a lagrangian structure.  In the superspace approach
it is a Super-Weil (SW) transformation \cite{GV} which relates
the system of constraints from \cite{GN2} (we define it as
set I) with that from \cite{N} (set
II).  Set I was  used in \cite{T1}. This set  produces the canonical
lagrangian. Set II was used in \cite{T2} and in the present paper.

 All quantities corresponding to  set I are
primed in the following to distinguish them from the same
objects in the set II. In all other respects we  follow closely
to the notations from \cite{T2}, \cite{T1}. We find the
SW-transformation in the form:

 $$ {E_a^M}' = \exp(2\rho)\,(E_a^M + f_a^\gamma\, E_\gamma^M), \ \ \ \
 {E_\beta^M}' = \exp(\rho) \, E_\beta^M      $$
 $$ {E_M^b}' = \exp(-2\rho)\, E_M^b, \ \ \ \
{E_M^\beta}' = \exp(-\rho)\, (E_M^\beta - E_M^b\, f_b^\beta)  $$
 $$ {D_\alpha}' = \exp(\rho) \, (D_\alpha +
 {1\over2}f_{\alpha,cd}{\hat M}^{cd}) $$
 $$ {D_a}' = \exp(2\rho)\, (D_a + f_a^\gamma\, D_\gamma +
 {1\over2}\, f_{a,cd}{\hat M}^{cd}),                    \eqno(6.1)  $$
where ${\hat M}^{cd}$ are $O(1.9)$-generators;
 $$  \rho = - {1\over 16} \log \phi, \ \ \rho_\beta \equiv D_\beta \, \rho =
 -( 16\phi)^{-1}\chi_\beta,
\ \  \ \ \rho_a \equiv D_a \rho                   $$
 $$ f_a^\gamma = -2 \Gamma_a^{\gamma \beta} \rho_\beta, \ \ \ \
 f_{\beta ab} = -4 {(\Gamma_{ab})_\beta}^\gamma \rho_\gamma, $$
 $$ f_{[a,b]c} = T_{abc} + \Sigma_{abc} + 4\eta_{c[a}\rho_{b]}, \ \ \ \
 {f^a}_{,ab} = -36\rho_b                            \eqno(6.2) $$
 where
$$ \Sigma_{abc} \equiv \rho\Gamma_{abc}\rho = ( 256
 \phi^2)^{-1} \, s_{abc}, $$
 but $ s_{abc} \equiv \chi \Gamma_{abc} \chi $.

 The relations (6.2) may be derived  if one calculates
 the  primed torsion-components:
 $$ {{T_{BC}}^A}' \equiv (-1)^{b(m+c)}{E_C^M}'{E_B^N}'{{T_{NM}}^A}' =$$
$$= (-1)^{b(n+c)}{E_C^N}'{D_B}'{E_N^A}'
  - (-1)^{cn}{E_B^N}'{D_C}'{E_N^A}'   $$
 in terms of the unprimed torsion-components using (6.1).
 By this way one obtains
the equations which may be solved immediately,
because ${T_{AB}^C}' $ and $T_{AB}^C$  are known from the
solution of corresponding BI's.

 The same procedure
 may be applied  to the $ {N_{A_1\ldots A_7}}' $.  (One must  take into
 account the factor $-1/2$ due to the different normalization of
the graviphoton field in the notations of \cite{T2} and \cite{T1}). Note
 that the SW transformation (6.1) does not affect world-space components,
 so:
$$N_{M_1\ldots M_7}' = -{1\over2} N_{M_1\ldots M_7}   $$
$$ {\cal F}_{MN}' = {\cal F}_{MN}  \eqno(6.3)  $$

 By this  way one also finds  the relation between primed and unprimed
 sets of physical fields. The result is:
 $$ e_m^a = \exp({1\over6}\phi')\, {e_m^a}' $$
 $$   \phi = \exp( - {4\over3}\phi') , \ \ \ \ \chi =
  -{4\over3}\exp(-{17\over12} \phi')\, \chi'  $$
 $$ \psi_m = \exp({1\over 12}\phi')({\psi_m}' - {1\over6}{\Gamma_m}'\chi') $$
 $$ N_{abc} = -2\exp(-{7\over6}\phi')\, L_{abc}'-
  {7\over12}\exp(-{1\over6}\phi')\, s_{abc}' $$
 $$ A_{abc} = {1\over3}\exp(-{3\over2}\phi')\, Z_{abc}',
  $$
$$ F_{ab} = \exp(-{\phi'\over 3})\, F_{ab}', \ \ \
\lambda = \exp(-{\phi'\over 4})\, \lambda' \eqno(6.4) $$
where $L', Z', s' $ are defined in \cite{T1} (these
objects appear in \cite{T1} without primes!),
 ${E_m^\alpha}' = {\psi_m^\alpha}' $ and $\Gamma_m'=
  {e_m^a}' \Gamma_a \  $, $\lambda'$ is defined according to (2.4) in terms
of $ ({{\cal F}^a}_\beta)'$.

To be complete we present also the kinetic part of the lagrangian
$ {\cal L}_{tot}'$\ \ $ (e{\cal L}_{tot} = e'\,{\cal L}_{tot}')$:

$$ {\cal L}_{tot}' = {1\over g^2} \left( -{1\over4}\,\exp(\phi')\,(F_{ab}')^2
+ \exp(\phi')\, \lambda'{\hat \nabla}'\lambda' \right) +
 R' + 2\,(\partial_a'\phi')^2 - $$
 $$ -{1\over3}\, \exp(-2\phi')\,
(M_{abc}')^2 - 2\,\psi_a'\Gamma^{abc}\nabla_b'\psi_c'+
4\, \chi'\nabla'\chi'         \eqno(6.5) $$

Note also relations, connecting matrix elements in the 16-component
formalism used here with the corresponding quantities in the
32-component formalism:

$$ \psi \Gamma_{(2k+1)}\psi = i {\bar \Psi} \gamma_{(2k+1)} \Psi, \ \ \
(\mbox{the same for $lambda$ and $\Lambda$}) $$
$$   \chi \Gamma_{(2k+1)}\chi = - i {\bar
 X} \gamma_{(2k+1)} X,   $$
$$ \psi \Gamma_{(2k)}\chi = i {\bar \Psi} \gamma_{(2k)}X,\ \ \
\chi \Gamma_{(2k)}\psi = -i {\bar X} \gamma_{(2k+1)} \Psi \, ,\eqno(6.6)$$
where $\Psi, \Lambda, X, \gamma_{n} \  $ are the 32-component formalizm
analogs of $\psi, \lambda, \chi, \Gamma_{n} $. (Note, that $\Psi, \Lambda$
 are 32-component spinors with positive, but $X$ -
with negative chirality). Eq.'s (6.4)-(6.6) provide the complete correspondence
between our
notations and that from other papers.

\section{Scaling Transformation}

The D=10 supergravity equations of motion are
invariant under the scale transformation of the type \cite{W}, \cite{GN2}:

   $$ X_j \rightarrow \mu^{q_j} \, X_j         \eqno(7.1) $$
where $X_j$ is an arbitrary  field, but $q_j$ is a numerical
factor, which has a specific value for each field, $\mu$ is an
arbitrary common factor.  This invariance may be reproduced at
the lagrangian level if one transforms  a lagrangian according
to the general rule (7.1) with $q=3$.  (The transformation (7.1)
does not touch the space-time coordinates).

 It is important that this invariance also takes place when
matter fields and tree-level superstring corrections are taken
into account, i.e.  equations of motions (A.6)-(A.11) and
equations (4.6) -(4.10) are scale-invariant.
 (Note that corrections of higher order in
the string-slope $\alpha'$ -parameter as well as one-loop
supergravity corrections break this invariance).

  We present  below the transformation rules for
different fields  (the numerical factors in the table are the
values of $q_j$ for each field):
  $$
\begin{array}{|c|c||c|c||c|c|} \hline
\phi & -1       & T_{abc}& -{1/2} &T_{ab}^\gamma & -{3/4} \\  \hline
 e_m^a &{1/2} &H_{abc}& -{3/2} &\psi_a^\gamma & - {1/4} \\ \hline
 D_a & -{1/2} & N_{abc} & -{1/2} & \chi    & -{5/4}       \\ \hline
D_\alpha & -{1/4} &A_{abc}& -{3/2}& {R_{ab}}^{cd} & -1 \\ \hline
{\cal F}_{ab} & -1 & \lambda & -{3/4} & e^{-1} {\cal L} & -2  \\ \hline
\end{array} $$

 This scale invariance is extremely helpful in establishing of
the lagrangian general structure  and the structure of any
possible intermediate expression.  It is this invariance helps
us to select in \cite{T2} the tree-level superstring/fivebrane
corrections from all other possible superstring correction
 terms in the equations of
motion. It is also the basis for use in
\cite{T2} the simplest  form of the  $N$-field BI's (with
zero in the r.h.s), as opposed to the case of $H$-field BI's in
the usual supergravity, where Chern-Simons contributions enter
in the r.h.s. (There is no possibillity to introduce the
Chern-Simons form into the r.h.s. of the $N$-field BI's
 without breaking
the (7.1) scale invariance, as opposed to the case of $H$-field
BI's). The corrections, related with the Green-Schwarz anomaly
compensating terms enter in the game only through the
$A_{abc}$-field in the DUAL SUGRA and do not break the (7.1) scale
invariance.

\section*{Appendix}

We present here  relations between the superfields
(and their zero superspace components) in the DUAL SUGRA, which
were derived in \cite{T2} and used in the text of the present
paper.

The torsion and the graviphoton BI's are used in the form:

$$ D_{[A}{T_{BC)}}^D + {T_{[AB}}^Q\, {T_{QC)}}^D  -
{{\cal R}_{[ABC)}}^D  = 0.                 \eqno(A.1) $$

$$ D_{[A_1}N_{A_2 \ldots A_8)} + {7\over2}\, {T_{[A_1A_2}}^Q \,
N_{QA_3\ldots A_8)} \equiv 0                                  \eqno(A.2) $$

The nonzero superfield components are
 $T_{abc}$ ( which is completely antisymmetric),
${T_{ab}}^\beta$ and:
$$ {T_{\alpha\beta}}^c = \Gamma_{\alpha\beta}^c \, , \ \ \
 {T_{a\beta}}^\gamma ={1\over 72}{({\hat T}\Gamma_a)_\beta}^\gamma \, ,$$
$$ N_{\alpha\beta a_1 \ldots a_5} =
 - (\Gamma_{a_1\ldots a_5})_{\alpha\beta} ,   $$
    $$  N_{abc} = T_{abc} \, ,
                                                            \eqno(A.3) $$
 where
$$ N_{abc} \equiv {1\over 7!} {\epsilon_{abc}}^{b_1\ldots b_7}
\, N_{b_1\ldots b_7}\, \ \ \ {\hat T} \equiv T_{abc}\Gamma^{abc}.   $$

All the   super-curvature components  are not equal to zero and may be
derived  in terms of torsion components and their spinorial
derivatives.
  There are constraints:

$$ D^aT_{abc} = 0, \eqno(A.4a)  $$
$$T_{ab}\Gamma^{ab} =0, \eqno(A.4b) $$
 $$  {\cal R} - {1\over 3}\, T^2 =0,  \eqno(A.4c)  $$
where ${\cal R}$ is a supercurvature scalar
 (${\cal R} \equiv {\cal R}_{abcd}\eta^{ac}\eta^{bd}$, \ \
 \   $T^2 \equiv T_{abc}T^{abc}$).
(There are a lot of additional relations, which are not interesting for
our purposes here, see \cite{T2} for details).

The dilaton $\phi$ and dilatino $\chi_\alpha \equiv D_\alpha
\phi$- superfields are introduced independently. The
$A_{abc}$-superfield (which is an arbitrary field up to the moment)
appears for the first time in the most general expresion of the
dilatino-field spinorial derivative:

 $$ D_\alpha \chi_\beta = -{1\over2}{\hat D}_{\alpha \beta}\phi
 +(-{1\over 36} \phi T_{abc} + A_{abc})\, \Gamma^{abc}_{\alpha \beta},
                                                            \eqno(A.5) $$

Now we are ready to present the complete set of e.m.'s
for the independent superfields.  (For our present purposes it
is sufficient to consider only zero superspace components of
these equations):\\
gravitino equation of motion:
     $$ Q_a \equiv \phi L_a - D_a\chi - {1\over 36} \Gamma_a {\hat
T}\chi - {1\over 24} {\hat T}\Gamma_a\chi + {1\over 42} \Gamma_a
\Gamma^{ijk}DA_{ijk} + {1\over 7} \Gamma^{ijk}\Gamma_a DA_{ijk} = 0,
                                                            \eqno(A.6) $$
dilatino equation of motion:
$$ Q \equiv {\hat D}\chi + {1\over 9}{\hat T}\chi +
 {1\over 3}\Gamma^{ijk}DA_{ijk} = 0.
                                                              \eqno(A.7)  $$
dilaton equation of motion:
$$ S \equiv D_a^2 \phi + {1\over 18}\phi T^2 - 2\, TA -
  {1\over 24} D\Gamma^{ijk}DA_{ijk} = 0.
                                                             \eqno(A.8)  $$
graviton equation of motion:
 $$ S_{ab} \equiv \phi {\cal R}_{ab} -  L_{(a}\Gamma_{b)}\chi -
 {1\over 36}\phi\eta_{ab}T^2 +
 D_{(a}D_{b)}\phi -$$
 $$-2\, T_{(a}A_{b)} + {3\over 28}D{\Gamma^{ij}}_{(a}DA_{b)ij} -
 {5\over 336}\eta_{ab} D\Gamma^{ijk}DA_{ijk}=0.
                                                            \eqno(A.9)$$
graviphoton equation of motion:
$$S_{abcd} \equiv D_{[a}(\phi T_{bcd]}) + {3\over 2} T_{[ab}\Gamma_{cd]}\chi
 + {3\over 2}
\phi T^2_{[abcd]} +$$
 $$+ {1\over 12} (T\epsilon A)_{abcd} + 6\,  (TA)_{[abcd]}
 + {3\over 4}D{\Gamma_{[ab}}^jDA_{cd]j} = 0.
                                                           \eqno(A.10) $$
The following notations are introduced in (A.6)-(A.10):
$$TA =
T_{ijk}A^{ijk}, \ \  (TA)_{ab} = T_{aij}{A_b}^{ij},  \ \ L_a = T_{ab}\Gamma^b
   $$
$$   (TA)_{abcd} = T_{abj}
{A_{cd}}^j, \ \ (T\epsilon A)_{abcd} = T^{ijk}\varepsilon_{ijkabcdmns}
A^{mns}.   $$
There are two  additional equations for the $A_{abc}$-superfield. The
first one follows from the self-consistency of  eq.(A.6) (cf.
\cite{N}, \cite{T1}, \cite{T3}):

$$ {D\Gamma_{[a}}^{ij}D\, A_{b]ij} + 56\, D^jA_{jab}
- {64\over 3} (TA)_{[ab]} =0.
                                                         \eqno(A.11) $$

The second one means, that 1200 IR contribution to the $A$-field
spinorial derivative is equal to zero:

   $$ (D_\alpha A_{abc})^{(1200)} = 0,
                                                            \eqno(A.12) $$

 This condition  may be derived immediately from (A.5)
\cite{T3}. Note, that the most general solution of (A.12) takes
the form:

$$ DA_{abc} = {\Gamma_{abc}}^{de} \, X_{de}.              \eqno(A.13) $$
where $ X_{ab}^\gamma $  is an arbitrary function which is $16+144+560$ IR of
O(1.9).

Using (A.13) one may may get rid of spinorial derivatives in the
equations of motion and consider them as equations for zero
superfield components. The explicit expression of
$X_{ab}$-superfield in terms of physical fields may be derived
using  (2.5) and (4.4) (for matter sector contribution) and
using  eq. (3.19) from
\cite{T2} (for superstring corrections contribution).

\end{document}